\documentclass[preprint,aps,nofootinbib]{revtex4}
\usepackage{graphicx,subfig}
\usepackage{dcolumn}
\usepackage{amsfonts, amssymb,amsmath}
\usepackage{hhline}
\usepackage{multirow}
\usepackage{bm}
 
\begin{document}
%$\tau \ \ \mathbf{\tau}\ \ \boldsymbol{\tau}\ \ \mathbf{g}$
\title{Are the chiral based $\bar{K}N$ potentials really energy-dependent?}

\author{J. R\'{e}vai}
\affiliation{Wigner Research Center for Physics, RMI,
H-1525 Budapest, P.O.B. 49, Hungary}
\date{\today}

\begin{abstract}
It is shown, that the energy dependence of the chiral based $\bar{K}N$ potentials, responsible for the occurrence of two poles in the $I=0$ sector, is the 
consequence of applying the on-shell factorization introduced in [E. Oset, A. Ramos, Nucl. Phys. A 635(1998)99]. When the dynamical equation is solved 
without this approximation, the $T$-matrix has only one pole in the energy region of the $\Lambda(1405)$ resonance.

\end{abstract}
\maketitle

\section{Introduction}

In the last two decades the $\bar{K}N$ interaction has attracted considerable attention as a basic input for studying the possible existence of $\bar{K}$-nuckear
clusters or states. The most challenging problem in this field is the structure of the $\Lambda(1405)$ resonance, and in particular, whether the observed bump in the 
$\pi\Sigma$ invariant mass distribution -- identified by the PDG as the  $\Lambda(1405)$ -- corresponds to a single pole of the scattering amplitude or is it a result of the interplay of two poles. We do not wish to recapitulate here the abundant literature in support of each opinion, a comprehensive list of  references can be 
found e.g. in the review paper \cite{list}. For our purpose it is enough to state, that according to the overall accepted opinion, the (multichannel) $\bar{K}N$
interaction, derived from the $SU(3)$ chiral perturbation theory uniquely supports the two-pole picture. It has to be noted, that there is still no consensus about the experimental situation concerning the pole struce of $ \Lambda(1405)$  \cite{Magas, AY1}.
The last and most accurate observation of the $\Lambda(1405)$ was performed in the CLAS $\gamma + p\rightarrow K^{+} + \Sigma +\pi$ photoproduction experiment \cite{CLAS1}. The obtained data were analyzed in several successive steps both by the participants of the experiment \cite{CLAS2,CLAS3,CLAS4} and by several theoretical papers  \cite{RO,MM}. In spite of the claim \cite{MM}, that the line-shape data definitely confirm the two-pole structure of the $\Lambda(1405)$, we think, that the presented proofs of this statement are not convincing:
\begin{itemize}
\item one-pole fits to the  $I=0\ \  \Sigma^{0}\pi^{0}$ data are at least as satisfactory as  the two-pole ones \cite{CLAS2,AY2};
\item a priori assuming the existence of two poles, the different fits lead to completely different pole positions, e.g. the "forced" two-pole positions in
    \cite{CLAS4} and \cite{MM} have nothing to do with each other. 
\end{itemize}
The aim of the present paper, however, is not to scrutinize different conclusions drawn from the CLAS data, but to draw attention to a common inconsistency in the derivation of the $\bar{K}N$ potential within the chiral approach. We shall present arguments in favor of the one-pole scenario, still compatible with the popular chiral based $\bar{K}N$ interactions. Our considerations will be performed for the lowest order Weinberg-Tomozawa (WT) term of the chiral interaction, since this is the only one from which one can derive a potential, suitable for systems involving a $K^-$-meson and more than one nucleon. Having in mind these calculations, the framework, we shall use, is non-relativistic quantum mechanics, with Lippmann-Schwinger (LS)-type basic dynamical equation. For regularization of divergent 
integrals in it, we shall use a separable potential model with suitable cut-off functions. 

In the above specified framework several successful calculations were performed for low-energy $K^-$-nuclear systems: in the two-body sector, mainly to adjust the potentials to the available experimental data \cite{CS,isospin,IKS}, for three-body systems using Faddeev-equations in search of $(\bar{K}pp)$ or $(\bar{K}\bar{K}p)$ quasi-bound states, see e.g. \cite{Shev,IS} and for $(K^-d)$ scattering \cite{RJFBS,Ohn} , 
and even for four-body systems, based on the corresponding AGS or Faddeev-Yakubovski equations \cite{Marri}. We think, that while in the two-body case the "full relativization" of this approach could lead to slightly different results, mainly compensated by parameter adjustment, for $n>2$ systems at present it seems to be inconceivable.

\section{The full WT potential $\bm{V}$ }

Our starting point is the lowest order Weinberg-Tomozawa term of the chiral Lagrangian ( eq.(7) from the basic paper \cite{Oset_Ramos}):
\begin {equation}
\label{orig}
\langle q_i |v_{ij}|q_j\rangle \sim - {c_{ij} \over 4 f_{\pi}^2}(q_i^0+q_j^0),
\end{equation}
where $i$ and $j$ denote the different meson-baryon channels, $c_{ij}$ are the SU(3) Clebsch-Gordan coefficients (tabulated in isospin basis e.g. in \cite{Hyodo_Weise}), $f_{\pi}$ is the pion decay constant, $q_i$ and $q_i^0=\sqrt{m_i^2+q_i^2}$ denote the meson c.m. momentum and energy
 in channel  $i$. We shall use isospin basis states, ($i=1,2,3,4,5$) correspond to ($[\bar{K}N]^{I=0},[\bar{K}N]^{I=1},[\pi\Sigma]^{I=0},[\pi\Sigma]^{I=1},
 [\pi\Lambda]^{I=1}$), respectively. When this expression is used as a kernel operator in dynamical equations, it has to be supplemented by normalization factors to yield properly normalized physical quantities as $T$-matrices, scattering amplitudes and cross sections. The actual form of these factors can depend on the type of the dynamical equation : Bethe-Salpeter or LS, on the treatment of kinematics : relativistic or non-relativistic or even semi relativistic (with reduced mass replaced by reduced energy) and on the normalization of the basis states. In this paper we shall use LS equation and non-relativistic kinematics. For this case the (s-wave)  potential (\ref{orig}) has the form:
 \begin{equation}
 \label{vnorm}
 \langle q_i|v_{ij}|q_j\rangle=-{c_{ij}\over 64 \pi^3 F_iF_j}\sqrt{{m_i+M_i\over m_i W}}\sqrt{{m_j+M_j\over m_j W}}({q_i^0}'+{q_j^0}'),
 \end{equation}
 where $m_i$ and $M_i$ are the meson and baryon masses in channel $i$, and $W$ is the total c.m. energy. In (\ref{vnorm}) 
 apart from the appropriate normalization factors, two commonly used modifications were introduced: the original meson decay constant $f_\pi$ was replaced by channel-dependent meson decay constants $F_i$ ($i=\pi,K$) and the meson energies in the last factor were modified by a relativistic correction (in fact by the baryon kinetic energy)\cite{KSW,CS}:
 \begin{equation}
  {q_i^0}'=q_i^0+{{q_i^0}^2-m_i^2\over 2 M_i}=q_i^0+{q_i^2\over 2 M_i}\,\, {\underset{\rm nonrel}\approx}\,\,m_i+{q_i^2\over 2 \mu_i}
  \end{equation}
  with the reduced mass $\mu_i=m_i M_i/(m_i+M_i)$.
  To calculate the $T$-matrix corresponding to the potential (\ref{vnorm}) we have to solve the LS equation. In order to ensure the convergence of the 
  occurring integrals a regularization method has to be applied. We use the separable potential representation of the $\bar{K}N$ interaction, which amounts to 
  multiplying the potential (\ref{vnorm}) by suitable cut-off factors $u_i(q_i)$ and $u_j(q_j)$ to obtain our final - somewhat unusual -  multichannel, two-term
  separable potential
   \begin{equation}
 \label{vfinal}
  \langle q_i|V_{ij}|q_j\rangle=u_i(q_i) \langle q_i|v_{ij}|q_j\rangle u_j(q_j)= \lambda_{ij}(g_{iA}(q_i)g_{jB}(q_j)+g_{iB}(q_i)g_{jA}(q_j))
 \end{equation} 
 with
$$
% \begin{equation}
  g_{iA}(q_i)=u_i(q_i); \qquad g_{iB}(q_i)=g_{iA}(q_i)\gamma_i(q_i)=g_{iA}(q_i)(m_i+{q_i^2\over 2\mu_i}).
$$
%\end{equation}
and
$$
 \lambda_{ij}=-{c_{ij}\over 64 \pi^3 F_iF_j}\sqrt{{m_i+M_i\over m_i W}}\sqrt{{m_j+M_j\over m_j W}}
$$ 

The LS equation for the $T$-matrix reads:
\begin{equation}
\label{LS}
\langle q_i|T_{ij}(W)|q_j\rangle =\langle q_i|V_{ij}|q_j\rangle+\sum_s\int \langle q_i|V_{is}|q_s\rangle G_s(q_s;W)\langle q_s|T_{sj}(W)|q_j\rangle d\vec{q}_s
\end{equation}
with the non-relativistic propagator
\begin{equation}
\label{prop}
G_s(q_s;W)=(W-m_s-M_s-{q_s^2\over 2 \mu_s}+i\epsilon)^{-1}={2\mu_s\over k_s^2-q_s^2+i\epsilon },
\end{equation}
where $k_s=\sqrt{2\mu_s (W-m_s-M_s)}$ is the on-shell c.m. momentum in channel $s$.

A commonly used procedure before solving the integral equation (\ref{LS}) is to remove the ``inherent'' $q$-dependence of the potential by replacing 
$q_i$ in $\gamma_i(q_i)$ by its on-shell value $k_i$
\begin{equation}
\gamma_i(q_i)\rightarrow \gamma_i(k_i)=W-M_i
\end{equation}
 This is the so-called on-shell factorization, introduced in \cite{Oset_Ramos} and never checked afterwards. 
Using this approximation the last factor in (\ref{vnorm}) is transformed into the familiar energy-dependent factor $(2W-M_i-M_j)$, responsible e.g.
for the appearance of a second pole in the $\bar{K}N-\pi\Sigma$ system. The form of the potential (\ref{vfinal}) offers a possibility to check the
validity/applicability of the on-shell factorization. The solution - obtaining a T-matrix via the solution of the LS equation - is straightforward, involving 
$2n\times 2n$ matrices, instead of the $n\times n$ ones, occurring in the one-term case ($n$ is the number of channels). 

Introducing the concise matrix notations
\begin{equation}
|\bm{q}\rangle=\begin{pmatrix}
|q_1\rangle&\hdots&0\\
\vdots&\ddots&\vdots\\
0&\hdots&|q_n\rangle
\end{pmatrix},\,\,|\bm{g_A}\rangle=\begin{pmatrix}
|g_{1A}\rangle&\hdots&0\\
\vdots&\ddots&\vdots\\
0&\hdots&|g_{nA}\rangle
\end{pmatrix},\,\,|\bm{g_B}\rangle=\begin{pmatrix}
|g_{1B}\rangle&\hdots&0\\
\vdots&\ddots&\vdots\\
0&\hdots&|g_{nB}\rangle
\end{pmatrix}
\end{equation}
and 
\begin{equation}
\bm{G}(W)=\begin{pmatrix}
G_1(q_1,W)&\hdots&0\\
\vdots&\ddots&\vdots\\
0&\hdots&G_n(q_n,W)
\end{pmatrix}
\end{equation}
eq.(\ref{vfinal}) can be written as $(\langle\bm{q}|\bm{V}|\bm{q}\rangle)_{ij}$ with 
\begin{equation}
\label{vecV}
\bm{V}=|\bm{g_A}\rangle\bm{\lambda}\langle\bm{g_B}|+|\bm{g_B}\rangle\bm{\lambda}\langle\bm{g_A}|
\end{equation}
Here and in the following bold face letters denote $n\times n$ matrices. 

The $T$-matrix has the form
\begin{equation}
\bm{T}=|\bm{g_A}\rangle\bm{\tau_{AA}}\langle\bm{g_A}|+|\bm{g_A}\rangle\bm{\tau_{AB}}\langle\bm{g_B}|+
|\bm{g_B}\rangle\bm{\tau_{BA}}\langle\bm{g_A}|+|\bm{g_B}\rangle\bm{\tau_{BB}}\langle\bm{g_B}|,
\end{equation}
where the $\bm{\tau}$ matrices are $n\times n$ sub-matrices of the $2n\times 2n$ matrix $M^{-1}$:
\begin{equation}
\label{inv}
M^{-1}=\begin{pmatrix}
\bm{\lambda}^{-1}-\langle\bm{g_B}|\bm{G}|\bm{g_A}\rangle&-\langle\bm{g_B}|\bm{G}|\bm{g_B}\rangle\\
-\langle\bm{g_A}|\bm{G}|\bm{g_A}\rangle&\bm{\lambda}^{-1}-\langle\bm{g_A}|\bm{G}|\bm{g_B}\rangle
\end{pmatrix}^{-1}=\begin{pmatrix}
\bm{\tau_{AB}}&\bm{\tau_{AA}}\\
\bm{\tau_{BB}}&\bm{\tau_{BA}}
\end{pmatrix}
\end{equation}

Due to the $q^2$ term in $\gamma_i(q)$, the requirement of convergence of all  Green's function matrix elements in (\ref{inv}), rules out the most commonly
used Yamaguchi form-factors; as a minimal extension, we took its square, the s.c. dipole form-factor:
\begin{equation}
u_i(q)=\left({\beta_i^2\over q^2+\beta_i^2}\right)^2,
\end{equation}
which was already used in the context of $\bar{K}N$ interactions \cite{IKS,IS}.

Let us consider one of these matrix elements, containing $|\bm{g_B}\rangle$:
\begin{equation}
\label{gm}
(\bm{G_{BA}})_{ij}=(\langle\bm{g_B}|\bm{G}|\bm{g_A}\rangle)_{ij}=\delta_{ij} 2\mu_i\int{u_i(q)^2\gamma_i(q)\over k_i^2-q^2+i\epsilon}d\vec{q}
\end{equation}
On-shell factorization means, that in (\ref{gm}) $\gamma_i(q)$ is replaced by $\gamma_i(k_i)=W-M_i$ and taken out from the integral. Performing this operation in all matrix elements, the on-shell matrix elements of the $\bm{T}$ operator can be written as:
\begin{equation}
\langle\bm{k}|\bm{T}|\bm{k}\rangle=\langle\bm{k}|\bm{g_A}\rangle\bm{\tau}\langle\bm{g_A}|\bm{k}\rangle
\end{equation}
with
\begin{equation}
\label{tau}
\bm{\tau}=\bm{\tau_{AA}+\tau_{AB}\gamma+\gamma\tau_{BA}+\gamma\tau_{BB}\gamma}.
\end{equation}
Here we introduced the matrix 
\begin{equation}
\bm{\gamma}=\begin{pmatrix}
\gamma_1(k_1)&\hdots&0\\
\vdots&\ddots&\vdots\\
0&\hdots&\gamma_n(k_n)
\end{pmatrix}
\end{equation}
and used the fact, that on-shell we have $|\bm{g_B}\rangle=|\bm{g_A}\rangle\bm{\gamma}$.
Using some matrix algebra one can show, that the $\bm{\tau}$ matrix of eq.(\ref{tau}) can be written as 
\begin{equation}
\bm{\tau}=((\bm{\lambda}\bm{\gamma}+\bm{\gamma}\bm{\lambda})^{-1}-\bm{G_{AA}})^{-1},
\end{equation}
which coincides with the corresponding $n\times n$ matrix $\bm{\tau}$  of the on-shell factorized potential
\begin{equation}
\label{vecU}
\bm{U}=|\bm{g_A}\rangle(\bm{\lambda}\bm{\gamma}+\bm{\gamma}\bm{\lambda})\langle\bm{g_A}|
\end{equation}
Now we are in the position to compare the results calculated from the ``full'' WT potential $\bm{V}$ of (\ref{vecV}) and its energy dependent counterpart
$\bm{U}$ of (\ref{vecU}), obtained by on-shell factorization. Both potentials depend on the same set of adjustable parameters: the two meson decay constants
$F_{\pi}$ and $F_K$ and the five cut-off ranges $\beta_i$.

Before proceeding to the discussion of detailed fits to the available experimental data by the two potentials, we make our most important statement:
for any reasonable combination of the parameters the full WT potential $\bm{V}$ produces only one pole below and close to the $\bar{K}N$ threshold, which can be associated with the $\Lambda(1405)$, while its counterpart $\bm{U}$ produces the familiar two poles: one close to the threshold and a second one much lower and broader. Thus it seems, that the occurrence of the second pole is not an ``inherent'' property of the chiral $SU(3)$ interactions (at least not of their WT term), but 
solely the consequence of applying the on-shell factorization approximation/method. It can be seen from eq.(\ref{gm}), that for real positive on-shell channel momenta $k_i$,
when the integrand is singular, the approximation might have some justification, e. g. for this case its imaginary part is reproduced exactly.
However, for investigating the analytical properties of the \hbox{$T$-matrix}, such as pole positions or sub-threshold amplitudes it can be -- and is in fact -- quite misleading. 

\section{Results of data fitting} 

When performing the fitting of the parameters of the potentials $\bm{U}$ and $\bm{V}$ to the experimental data we made a slight change in the calculation of
the $T$-matrices, as compared to what is described in the previous section.
In order to include in the fitting the important information of the SIDDHARTA $1s$ level shift in kaonic hydrogen \cite{sid}, 
the Green's function matrix $\bm{G}$ was calculated with the inclusion of the Coulomb interaction in  the $K^-p$ (particle) channel and  -- to ensure right 
threshold positions - with physical masses in the $\bar{K}N$ channels.
 This isospin breaking leads to a non-diagonal $\bm{G}$ matrix in isospin basis and mixing of $I=0$ and $I=1$ channels.
 The details of this procedure can be found in\hbox{ \cite{isospin}}. As a result, we get exact $1s$ level shifts, without reference to the popular  modified Deser formula \cite{des}\footnote{It is less known, that a further improvement of this formula exists \cite{des1}, which in the cases, when the exact values
 can be calculated (two- and three-body systems), reproduces the level shifts considerably better, than the original one.} .
 
\begin{figure}
\includegraphics[width=\textwidth]{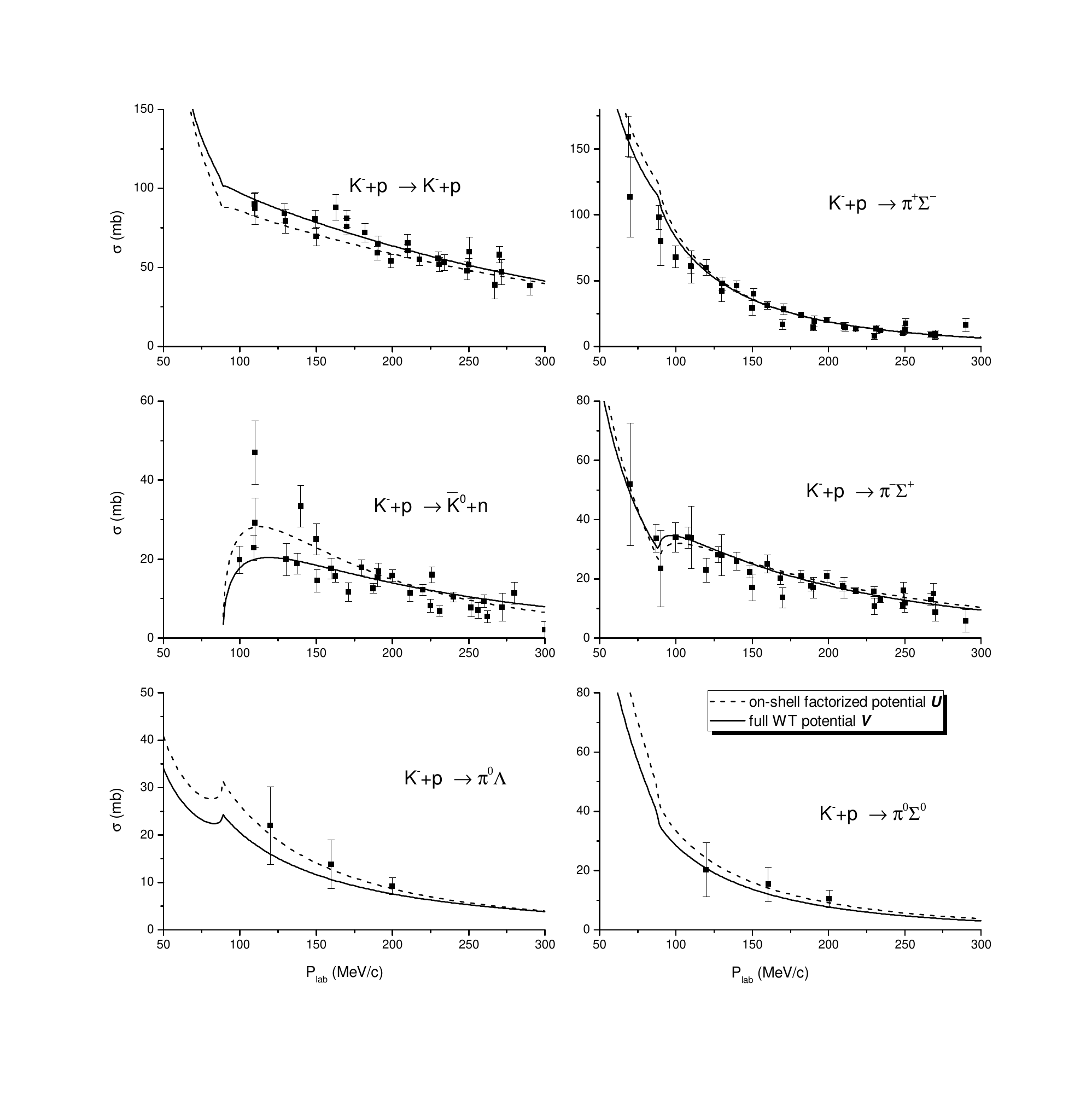}
\caption{Elastic and inelastic $K^-p$ cross sections for the potentials $\bm{U}$ and $\bm{V}$.}
\label{fig:cross}
\end{figure}

 In Fig.\ref{fig:cross} and Table \ref{disc} we show the results of fitting the potential parameters to experimental data. Apart from the usual six low-energy $K^-p$ cross sections we took four discrete observables to fix the parameters: the three threshold branching ratios
$$
 \gamma={\sigma(K^-p\rightarrow\Sigma^-\pi^+)\over \sigma(K^-p\rightarrow\Sigma^+\pi^-)};\ R_n={\sigma(K^-p\rightarrow \pi^0\Lambda)\over
 \sigma(K^-p\rightarrow \pi^0\Lambda, \pi^0\Sigma^0 )};\ 
 R_c={ \sigma(K^-p\rightarrow\Sigma^-\pi^+,\Sigma^+\pi^- )\over \sigma(K^-p\rightarrow {\rm all\ inelastic\ channels})}
$$ 
and the $1s$ level shift $\Delta E$ in kaonic hydrogen.

\begin{table*}
\begin{center}
\begin{tabular}{|c|c|c|c|}
\hline
&$\bm{V}$&$\bm{U}$&Exp\\
\hline
$\gamma$&$2.347$&$2.343$&$2.36\pm0.04$\\
$R_c$&$0.687$&$0.665$&$0.664\pm0.011$\\
$R_n$&$0.203$&$0.196$&$0.189\pm0.015$\\
$\Delta E\ (eV)$&$384-231\ i$&$313-285\ i$&$(283\pm36)-(271\pm46)\ i$\\
\hline
\end{tabular}
\end{center}
\caption{Calculated and experimental values of the discrete data for potentials  $\bm{U}$ and $\bm{V}$.}
\label{disc}
\end{table*}
 The parameters of the best fit potentials are shown in Table \ref{par}.
\begin{table*}
\begin{center}
\begin{tabular}{|c|c|c|c|c|c|c|c|}
\hline
&$F_{\pi}$&$F_K$&$\beta_1$&$\beta_2$&$\beta_3$&$\beta_4$&$\beta_5$\\
\hline
$\bm{V}$&73.16&98.29&830.2&934.6&451.8&471.2&352.4\\
$\bm{U}$&102.0&110.2&1270&1695&874.6&929.6&444.0\\
\hline
\end{tabular}
\end{center}
\caption{Parameters of the potentials $\bm{V}$ and $\bm{U}$ (all values are given in $MeV$).}
\label{par}
\end{table*}

The pole positions on the unphysical sheet of the $\pi\Sigma$ channel are
$$
z_1(\bm{U})=(1427-33i)\ MeV;\ \ z_2(\bm{U})=(1361-64i)\ MeV;\ \ z_1(\bm{V})=(1422-26i)\ MeV
$$

Probably, the obtained fits could be further improved, however, this was not the main aim of the present work. It can be seen, that equal quality fits can be 
achieved for both potentials $\bm{V}$ and $\bm{U}$. But the fact that the best fit parameters of the two potentials are quite different, shows that $\bm{U}$ can not be considered as an approximation to $\bm{V}$. These are different interactions: for the same parameter set they would  give completely different results. In this sense, 
the reasoning in \cite{Oset_Ramos}, according to which the on-shell factorization has to be accompanied by a readjustment of the parameters to reproduce the on-shell properties of the original interaction, seems to be correct. However, this is not true for the off-shell properties, such as pole structure and, maybe, others.

\section{Conclusions}

We have shown, that the energy dependence of the lowest order WT term of the $\bar{K}N$ interaction, derived from the chiral $SU(3)$ Lagrangian, follows 
from the on-shell factorization approximation, introduced in order to simplify the solution of the dynamical equation. Avoiding  this approximation an 
energy-independent $\bar{K}N$ potential $\bm{V}$ was derived, which supports only one pole in the region of the $\Lambda(1405)$ resonance. Thus the 
-- almost -- overall accepted view, that chiral based interactions lead to a ``two-pole structure of the $\Lambda(1405)$'', becomes questionable. The potential
$\bm{V}$, being energy-independent, is suitable for standard quantum mechanical calculations in $n>2$ systems, including coordinate space variational
approaches, where the energy-dependence leads to serious difficulties.  We think, that in view of these findings, a certain part of the huge work done 
in the field of strangeness nuclear physics, and in particular, concerning $\bar{K}$-nuclear clusters, has to be reconsidered.  
%\vspace{.5cm}

\noindent
{\bf Acknowledgment.}

The work was supported by the % Czech GACR grant P203/12/2126 and the
Hungarian OTKA grant 109462.

\end{document}